\documentclass[aps,prd,onecolumn,groupedaddress,showpacs,nofootinbib,amssymb]{revtex4}
\usepackage[dvips]{graphicx}
\usepackage{amssymb}
\usepackage{amsmath}
\usepackage{graphicx,,color}
\usepackage{amsfonts}
\usepackage{bm}
\usepackage{cancel}
\usepackage{comment}

\newcommand\be{\begin{equation}}
\newcommand\ee{\end{equation}}

\allowdisplaybreaks[4]

\begin{document}

\title{Pre-Inflationary Bounce Effects on Primordial Gravitational Waves of $f(R)$ Gravity}
\author{S.D. Odintsov,$^{1,2}$}
\email{odintsov@ice.cat}\author{V.K.
Oikonomou,$^{3,4}$}\email{v.k.oikonomou1979@gmail.com,voikonomou@auth.gr}
\affiliation{$^{1)}$ ICREA, Passeig Luis Companys, 23, 08010 Barcelona, Spain\\
$^{2)}$ Institute of Space Sciences (IEEC-CSIC) C. Can Magrans
s/n,
08193 Barcelona, Spain\\
$^{3)}$Department of Physics, Aristotle University of Thessaloniki, Thessaloniki 54124, Greece\\
$^{4)}$ Laboratory for Theoretical Cosmology, Tomsk State
University of Control Systems and Radioelectronics  (TUSUR),
634050 Tomsk, Russia}

\tolerance=5000

\begin{abstract}
In this work we shall study a possible pre-inflationary scenario
for our Universe and how this might be realized by $f(R)$ gravity.
Specifically, we shall introduce a scenario in which the Universe
in the pre-inflationary era contracts until it reaches a minimum
magnitude, and subsequently expands, slowly entering a slow-roll
quasi-de Sitter inflationary era. This pre-inflationary bounce
avoids the cosmic singularity, and for the eras before and after
the quasi-de Sitter inflationary stage, approximately satisfies
the string theory motivated scale factor duality
$a(t)=a^{-1}(-t)$. We investigate which approximate forms of
$f(R)$ can realize such a non-singular pre-inflationary scenario,
the quasi-de Sitter patch of which is described by an $R^2$
gravity, thus the exit from inflation is guaranteed. Furthermore,
since in string theory pre-Big Bang scenarios lead to an overall
amplification of the gravitational wave energy spectrum, we
examine in detail this perspective for the $f(R)$ gravity
generating this pre-inflationary non-singular bounce. As we show,
in the $f(R)$ gravity case, the energy spectrum of the primordial
gravitational waves background is also amplified, however the
drawback is that the amplification is too small to be detected by
future high frequency interferometers. Thus we conclude that, as
in the case of single scalar field theories, $f(R)$ gravity cannot
produce detectable stochastic gravitational waves and a
synergistic theory of scalars and higher order curvature terms
might be needed.
\end{abstract}


\maketitle

\section{Introduction}

Inflation is one of the most elegant theoretical proposals for
describing the classical evolving post-Planck era
\cite{inflation1,inflation2,inflation3,inflation4}, since it
solves most of theoretical inconsistencies of the standard Big
Bang cosmology. In the next two decades the inflationary scenario
will finally be scrutinized by stage four Cosmic Microwave
Background (CMB) experiments
\cite{CMB-S4:2016ple,SimonsObservatory:2019qwx}, and several high
frequency interferometers
\cite{Hild:2010id,Baker:2019nia,Smith:2019wny,Crowder:2005nr,Smith:2016jqs,Seto:2001qf,Kawamura:2020pcg,Bull:2018lat}.
These experiments will reveal either a $B$-mode polarization
pattern (curl modes) in the CMB \cite{Kamionkowski:2015yta} or
will detect a direct stochastic primordial gravitational wave
background. In all cases, the scientific community is anticipating
with great interest the results of these experiments.

Although quite appealing, the standard inflationary scenario
described by a single scalar field has some drawbacks, related
with the nature and interactions of the so-called inflaton field.
The mass, couplings, the potential energy, and several other
properties of the inflaton make quite difficult to identify this
particle in the context of Standard Model particle physics. This
inflaton is introduced by hand, and in an ad hoc way controls one
of the most important evolution eras of our Universe. However most
of the inflaton's properties are basically suitably fine-tuned. To
our opinion, scalar fields are expected to be present in the
inflationary Lagrangian, either alone or in the presence of higher
curvature terms. The scalar fields naturally appear in the most
successful to date theoretical description of high energy physics,
string theory in its various forms. Thus one would naturally
expect to see them present in the low energy effective
inflationary Lagrangian, where they can, in principle, control the
dynamics. But the low energy string theory inflationary Lagrangian
does not only contain scalar fields, but also higher order
curvature terms \cite{Codello:2015mba}. Hence the scalar fields
and the higher order curvature terms may synergistically control
the inflationary dynamics. In fact, it is possible to have higher
order curvature corrections in a minimally coupled scalar field
theory, or in a conformally coupled scalar field theory
\cite{Codello:2015mba}. Admittedly, it is much more simple to
perform calculations using a single scalar field, or some higher
order curvature modified gravity
\cite{reviews1,reviews2,reviews3,reviews4,reviews5,reviews6}. This
approach is much simpler and offers insights toward understanding
the inflationary era. To our opinion a synergistic approach that
combines scalars and higher order curvature terms, is the most
complete approach from many aspects, and there are many examples
of this sort in the literature
\cite{Ema:2017rqn,Ema:2020evi,Ivanov:2021ily,Gottlober:1993hp,delaCruz-Dombriz:2016bjj,Enckell:2018uic,Karam:2018mft,Kubo:2020fdd,Gorbunov:2018llf,Calmet:2016fsr,Oikonomou:2021msx}.
But as we said, studies of theories containing either scalar or
only higher order curvature terms offer a simple path toward
understanding the fundamental physics and problems of inflation.
In this paper we shall adopt this line of research and we shall
use the most appealing modified gravity theory, namely $f(R)$
gravity
\cite{Nojiri:2003ft,Capozziello:2005ku,Hwang:2001pu,Cognola:2005de,Song:2006ej,Faulkner:2006ub,Olmo:2006eh,Sawicki:2007tf,Faraoni:2007yn,Carloni:2007yv,
Nojiri:2007as,Deruelle:2007pt,Appleby:2008tv,Dunsby:2010wg}. We
shall introduce a new possibility for the era prior to the
inflationary era, in which the pre-inflationary era is described
by a non-singular bounce, followed by the inflationary era in the
form of a quasi-de Sitter era, which then may be followed by a
reheating era. Bouncing cosmology is an alternative scenario to
the standard inflationary scenario
\cite{Brandenberger:2012zb,Brandenberger:2016vhg,Battefeld:2014uga,Novello:2008ra,Cai:2014bea,deHaro:2015wda,Odintsov:2021yva,Banerjee:2020uil},
and here we shall study a pre-inflationary bounce cosmology
scenario. The inflationary dynamics is controlled geometrically by
a modified gravity in the form of a deformed $R^2$ model
\cite{Starobinsky:1980te,Bezrukov:2007ep}. More importantly during
the pre-inflationary era the Universe contracts until it reaches a
minimum size, avoiding the cosmic singularity. After that the
Universe inflates and the graceful exit is achieved by the $R^2$
gravity, which naturally contains unstable de Sitter attractors
\cite{Odintsov:2017tbc}. Using standard reconstruction techniques
\cite{Nojiri:2009kx} we shall investigate which $f(R)$ gravity can
approximately generate such a non-singular pre-inflationary epoch,
which leads to the standard quasi-de Sitter epoch described by
$R^2$ gravity. As we will show, the non-singular pre-inflationary
bounce approximately respects a string theory motivated duality,
the scale factor duality \cite{Veneziano:1991ek,Gasperini:2007vw},
and this symmetry is an exact symmetry all eras apart from the the
quasi-de Sitter epoch are considered. From string theory, it is
known that pre-inflationary scenarios lead to amplification of the
primordial gravitational wave energy spectrum
\cite{Gasperini:2007vw}, see also Refs.
\cite{Navascues:2021mxq,Anderson:2020hgg,Li:2019ipm,Cai:2015nya,Wang:2014abh,Kitazawa:2014dya,Rinaldi:2010yp}
for other pre-inflationary scenarios. As we will also show in our
case, this amplification occurs in $f(R)$ gravity too, in which
case, the pre-inflationary bounce amplifies the energy spectrum of
the primordial gravitational waves. However, as we show, the
amplification is very small, and this brings us to the argument
that even with this primordial epoch amplification, $f(R)$ gravity
alone, without the presence of a scalar field, will hardly
describe the physics of the primordial Universe alone, in the case
of a future detection of either the curl modes in the CMB or the
stochastic primordial gravitational wave background. Hence,
amplification in $f(R)$ gravity may occur if in addition to the
higher derivative terms, a scalar field (for example the Higgs
field) is present.

Before getting to the core of our analysis, let us mention that we
shall assume that the geometric background is a flat
Friedmann-Robertson-Walker (FRW) of the form,
\begin{equation}
\label{JGRG14} ds^2 = - dt^2 + a(t)^2 \sum_{i=1,2,3}
\left(dx^i\right)^2\, ,
\end{equation}
where $a(t)$ is the scale factor and also we shall adopt the
natural units physical system of units.

\section{Pre-Big Bag Bounce and Realization from $f(R)$ Gravity}

The inflationary era in standard cosmological contexts is
quantified by a de Sitter or a quasi-de Sitter evolution. In all
cases the inflationary era must last sufficiently long in order
for the problems of standard Big Bang cosmology to be resolved. In
most inflationary contexts, the Planck epoch precedes the
inflationary era, so basically the inflationary era is a classical
or semi-classical epoch in which most of the quantum effects of
the Planck epoch have been smoothed out, or at least have an
effective impact on the inflationary Lagrangian. In this section
we shall propose a pre-inflationary epoch which is described by a
non-singular bounce. We shall qualitatively analyze the physics of
this pre-inflationary epoch and  we shall calculate the $f(R)$
gravity which can realize such a non-singular bounce. Also we
shall discuss in detail how this pre-inflationary asymmetric
bounce shares an approximate symmetry common in string theories,
and specifically the scale factor duality
\cite{Veneziano:1991ek,Gasperini:2007vw}, which is a version of
the string theory T-duality in spacetimes with time-dependent
radius, such as the FRW spacetime. Interestingly enough, this
scale factor duality is violated only during the inflationary era,
and holds true before and after the inflationary era. In our
scenario, the scalar curvature perturbations relevant for the CMB
are generated during the inflationary era.

\begin{figure}[h!]
\centering
\includegraphics[width=16pc]{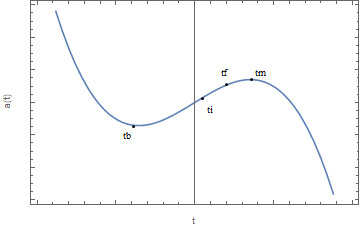}
\includegraphics[width=16pc]{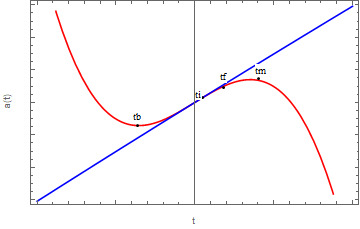}
\caption{The scale factor of the pre-inflationary non-singular
bounce evolution (left plot) and post-bounce inflationary part. In
the left with red curve we present the same pre-inflationary
non-singular bounce and with blue curve the pure quasi-de Sitter
part of the non-singular bounce. In both cases, the matter perfect
fluids are absent.} \label{plot1}
\end{figure}

\begin{figure}[h!]
\centering
\includegraphics[width=16pc]{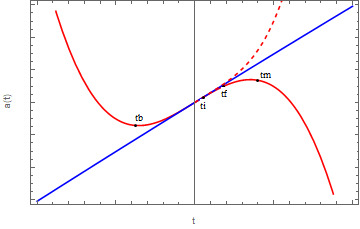}
\caption{The scale factor of the pre-inflationary non-singular
bounce evolution and post-bounce inflationary part (red curve),
the pure quasi-de Sitter part of the non-singular bounce (blue
curve) and the post-inflationary evolution (red dashed curve) in
the presence of perfect matter fluids post-inflationary.}
\label{plot2}
\end{figure}

Let us begin with the scale factor which describes a
pre-inflationary non-singular bounce and a subsequent quasi-de
Sitter era, which is,
\begin{equation}\label{scalefactorquasidesitter}
a(t)=a_0e^{-H_b^3t^3+H_0 t-H_i^2 t^2}\, ,
\end{equation}
where $a_0$ is the scale factor at the bouncing point time
instance $t_b$, thus the minimum scale that the Universe can
acquire. In Eq. (\ref{scalefactorquasidesitter}), the parameters
$H_b$, $H_0$ and $H_i$ have mass units $[H_b]=[H_0]=[H_i]=M$ in
natural units. The evolution corresponding to the scale factor
(\ref{scalefactorquasidesitter}) is an asymmetric bounce as it can
be seen in the left plot of Fig. \ref{plot1}. In the right plot of
Fig. \ref{plot1} we also present the simple quasi-de Sitter part
of the scale factor (\ref{scalefactorquasidesitter}), namely
$a(t)=a_0e^{H_0 t-H_i^2 t^2}$ (blue curve), and apparently before
$t=0$ and for negative times, the $t^3$ term starts to dominate
the evolution and eventually controls the dynamics. The scale
factor (\ref{scalefactorquasidesitter}) has a maximum and a
minimum, at two time instances, $t_b$ and $t_m$, which are,
$t_b=\frac{-\sqrt{3 H_0 H_b^3+H_i^4}-H_i^2}{3 H_b^3}$ and
$t_m=\frac{\sqrt{3 H_0 H_b^3+H_i^4}-H_i^2}{3 H_b^3}$, and clearly
$t_b<0$ while $t_m>0$.
Basically, $t_b$ is the bouncing point which occurs before $t<0$,
and $t_m$ would be the maximum of the scale factor after the
bounce, if radiation and dark matter fluids were neglected. The
physical scenario we propose is basically seen in Fig.
\ref{plot2}. In the complete scenario, the Universe evolves from
negative times, before $t_b$ in a contracting way, until it
bounces off at the time instance $t_b$. After $t=0$, at the time
instance $t_i$ the inflationary era commences which continues
until $t_f$, in such a way that 60 $e$-foldings are accomplished.
Between $t_f$ and $t_i$ the Universe is basically described by the
quasi-de Sitter evolution $a(t)=a_0e^{H_0 t-H_i^2 t^2}$ and beyond
that if radiation and matter fluids are ignored, the Universe's
scale factor would go to the maximum $t_m$ and then would decrease
in a cubic exponential way. However, after inflation, the matter
fluids cannot be ignored, therefore, the Universe will expand in a
decelerating way following the red dashed curve.
In this scenario, the curvature perturbations are generated during
the quasi-de Sitter stage, in which the Hubble radius decreases,
and the scalar quantum fluctuations are amplified thus providing
the seeds  for structure formation and the CMB temperature
fluctuations. Before the quasi-de Sitter era, no scalar
perturbations are produced, however it is interesting to note that
the pre-inflationary era might affect the energy power spectrum of
the primordial gravitational waves, giving an overall
amplification per-inflationary amplification factor to the general
relativistic waveform, as we show later on.

Interestingly enough, the scale factor
(\ref{scalefactorquasidesitter}) obeys an approximate string
originating duality, namely the scale factor duality
\cite{Veneziano:1991ek,Gasperini:2007vw}, which is,
\begin{equation}\label{scalefactorduality}
a(t)=a^{-1}(-t)\, .
\end{equation}
Indeed, when the $t^2$ term does not affect the evolution, the
scale factor duality is satisfied by the scale factor
(\ref{scalefactorquasidesitter}), so the symmetry
(\ref{scalefactorduality}) is respected during the post-bouncing
point and before the inflationary era, and after the inflationary
era. It is known \cite{Gasperini:2007vw} that string theories
which respect this scale factor duality lead to an overall
amplification of the primordial gravitational wave spectrum, and
we shall examine this perspective in the context of $f(R)$
gravity.

Let us now investigate how the scale factor
(\ref{scalefactorquasidesitter}) can be realized in the context of
vacuum $f(R)$ gravity. We shall be interested mainly in the
quasi-de Sitter regime and the pre-inflationary regime in which
the Universe evolves from the bouncing point until the beginning
of the inflationary era. In the former case, the scale factor is
approximately described by the quasi-de Sitter terms thus
$a(t)\sim^{H_0t-H_i^2t^2}$, while in the latter case $a(t)\sim
e^{-H_b^3t^3}$. In order to find which vacuum $f(R)$ gravity
realizes these evolution patches, we shall use a well-known
reconstruction technique developed in \cite{Nojiri:2009kx}. To
start with, we shall consider a vacuum $f(R)$ gravity theory with
the following gravitational action,
\begin{equation}\label{action1dse}
\mathcal{S}=\frac{1}{2\kappa^2}\int \mathrm{d}^4x\sqrt{-g}f(R),
\end{equation}
with $\kappa^2$ being as usual $\kappa^2=8\pi G=\frac{1}{M_p^2}$
and $M_p$ denotes the reduced Planck mass. The field equations are
found by varying the action with respect to the metric tensor,
\begin{equation}\label{eqnmotion}
f_R(R)R_{\mu \nu}(g)-\frac{1}{2}f(R)g_{\mu
\nu}-\nabla_{\mu}\nabla_{\nu}f_R(R)+g_{\mu \nu}\square f_R(R)=0\,
,
\end{equation}
where we introduced $f_R=\frac{\mathrm{d}f}{\mathrm{d}R}$. For the
FRW metric of Eq. (\ref{JGRG14}), the field equations of vacuum
$f(R)$ gravity take the following form,
\begin{align}
\label{JGRG15} 0 =& -\frac{f(R)}{2} + 3\left(H^2 + \dot H\right)
f_R(R) - 18 \left( 4H^2 \dot H + H \ddot H\right) f_{RR}(R)\, ,\\
\label{Cr4b} 0 =& \frac{f(R)}{2} - \left(\dot H +
3H^2\right)f_R(R) + 6 \left( 8H^2 \dot H + 4 {\dot H}^2 + 6 H
\ddot H + \dddot H\right) f_{RR}(R) + 36\left( 4H\dot H + \ddot
H\right)^2 f_{RRR} \, ,
\end{align}
where $f_{RR}=\frac{\mathrm{d}^2f}{\mathrm{d}R^2}$, and
$f_{RRR}=\frac{\mathrm{d}^3f}{\mathrm{d}R^3}$. Now we can apply
the reconstruction technique developed in  \cite{Nojiri:2009kx},
which is basically based on using the $e$-foldings number as a
dynamical variable,
\begin{equation}\label{efoldpoar}
e^{-N}=\frac{a_0}{a}\, ,
\end{equation}
where $a_0$ in our case is the scale factor of the Universe at the
bouncing point time instance $t_b$. Rewriting the Friedmann
equation (\ref{JGRG15}) in terms of $N$ we get,
\begin{equation}
\label{newfrw1} -18\left [ 4H^3(N)H'(N)+H^2(N)(H')^2+H^3(N)H''(N)
\right ]f_{RR}+3\left [H^2(N)+H(N)H'(N)
\right]f_R-\frac{f(R)}{2}=0,
\end{equation}
where the prime in the equation above, denotes differentiation
with respect to $N$. Upon introducing the function $G(N)=H^2(N)$,
the Ricci scalar takes the form,
\begin{equation}\label{riccinrelat}
R=3G'(N)+12G(N)\, ,
\end{equation}
and from the above equation, by inverting it, we may obtain
$N(R)$. In the case of the pure quasi-de Sitter evolution $a\sim
e^{H_0t-H_i^2t^2}$, we have,
\begin{equation}\label{gnfunction}
G(N)=H_0^2-4 H_i^2 N\, .
\end{equation}
By combining Eqs. (\ref{riccinrelat}) and (\ref{gnfunction}), the
$e$-foldings number $N$ as a function of the Ricci scalar reads,
\begin{equation}\label{efoldr}
N=\frac{12 H_0^2-12 H_i^2-R}{48 H_i^2}\, .
\end{equation}
Eventually, we can rewrite the Friedmann equation as a function
$G(N)$ as follows,
 \begin{equation}
\label{newfrw1modfrom} -9G(N(R))\left[ 4G'(N(R))+G''(N(R))
\right]F''(R) +\left[3G(N)+\frac{3}{2}G'(N(R))
\right]F'(R)-\frac{F(R)}{2}=0,
\end{equation}
with $G'(N)=\mathrm{d}G(N)/\mathrm{d}N$ and
$G''(N)=\mathrm{d}^2G(N)/\mathrm{d}N^2$. By using Eq.
(\ref{efoldr}), the Friedmann equation takes the following form,
\begin{align}
\label{bigdiffgeneral1} & \left(12 H_i^2 \left(12
H_i^2+R\right)\right)\frac{\mathrm{d}^2f(R)}{\mathrm{d}R^2} +
\left(\frac{R}{4}-3
H_i^2\right)R\frac{\mathrm{d}f(R)}{\mathrm{d}R}-\frac{f(R)}{2}=0,
\end{align}
which can be solved analytically and it yields the following
$f(R)$ gravity solution,
\begin{equation}\label{frformprev}
f(R)=R+\frac{R^2}{72 H_i^2}+2 H_i^2-\frac{\mathcal{C}_2 \left(144
H_i^4+72 H_i^2 R+R^2\right) \left(\sqrt[4]{e} \sqrt{3 \pi }
\text{erf}\left(\frac{\sqrt{4 H_i^2+\frac{R}{3}}}{4
H_i}\right)+\frac{12 H_i e^{-\frac{R}{48 H_i^2}} \left(36
H_i^2+R\right) \sqrt{12 H_i^2+R}}{144 H_i^4+72 H_i^2
R+R^2}\right)}{3981312 H_i^9} \, ,
\end{equation}
where $\mathcal{C}_2$ is an integration constant with mass
dimensions $[\mathcal{C}_2]=[M]^7$. The model (\ref{frformprev})
is a deformed $R^2$ model, and its phenomenology is proven to be
similar to the $R^2$ model (this will be shown elsewhere).
Specifically, for $N\sim 60$ it yields a spectral index of
primordial scalar curvature perturbations $n_s\sim 0.967078$,
irrespective of the values of the free parameters, which is very
close to the value of the $R^2$ model, $n_s=0.966667$.
Accordingly, the same applies for the tensor-to-scalar ratio and
the tensor spectral index, which are approximately $r=0.00327846$
and $n_T\simeq -0.000135483$ irrespective of the values of the
free parameters. The phenomenology of the model (\ref{frformprev})
is fully analyzed in another work by us, and it proves that the
model is a slight deformation of the standard $R^2$ model. Now let
us discuss which $f(R)$ gravity can realize the pre-inflationary
evolution, and specifically its dominant part, namely $a(t)\sim
e^{-H_b t^3}$. In this case we have,
\begin{equation}\label{gnfunctioncubic}
G(N)\simeq 9 H_b^2 N^{4/3}\, ,
\end{equation}
at leading order in the $e$-foldings number. Again, by combining
Eqs. (\ref{riccinrelat}) and (\ref{gnfunction}), the $e$-foldings
number $N$ as a function of the Ricci scalar reads,
\begin{equation}\label{efoldrcubic}
N=\frac{R^{3/4}}{18 \sqrt{2} \sqrt[4]{3} H_b^{3/2}}\, .
\end{equation}
By using Eq. (\ref{efoldr}), the Friedmann equation in this case,
takes the following form,
\begin{align}
\label{bigdiffgeneral1cubic} & \left(-6 \sqrt{2} \sqrt[4]{3}
H_b^{11/2} R^{5/4}\right)\frac{\mathrm{d}^2f(R)}{\mathrm{d}R^2}
+\frac{R}{4}\frac{\mathrm{d}f(R)}{\mathrm{d}R}-\frac{f(R)}{2}=0,
\end{align}
which can be solved analytically and it yields the following
$f(R)$ gravity solution,
\begin{equation}\label{frformprevcubic}
f(R)=\mathcal{C}_3\,
_1F_1\left(-\frac{8}{3};-\frac{1}{3};\frac{R^{3/4}}{18 \sqrt{2}
\sqrt[4]{3} H_b^{11/2}}\right)-\mathcal{C}_4 \frac{R \,
_1F_1\left(-\frac{4}{3};\frac{7}{3};\frac{R^{3/4}}{18 \sqrt{2}
\sqrt[4]{3} H_b^{11/2}}\right)}{108 H_b^{22/3}} \, ,
\end{equation}
where $\, _1F_1(a;b;z)$ is the Kummer confluent hypergeometric
function, and $\mathcal{C}_3$, $\mathcal{C}_4$ are integration
constants.

\begin{figure}[h!]
\centering
\includegraphics[width=40pc]{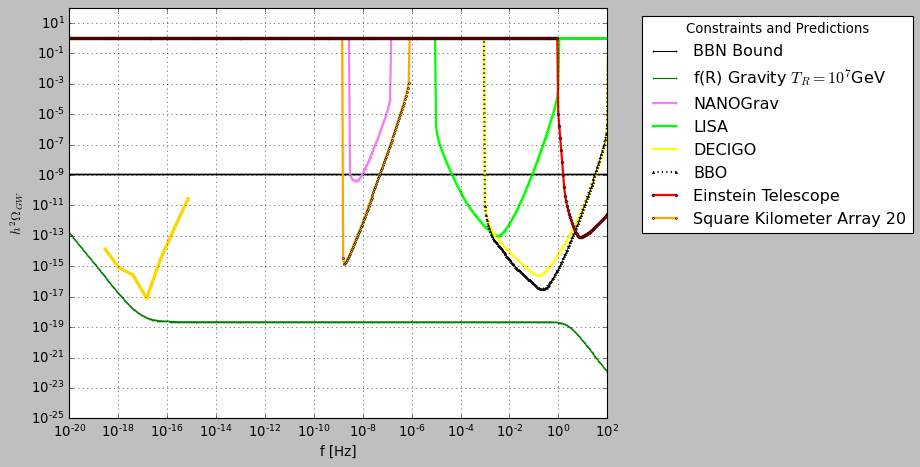}
\caption{The $h^2$-scaled gravitational wave energy spectrum for
pure $f(R)$ gravity.} \label{plotfinalfrpure}
\end{figure}

Having the $f(R)$ model which realizes the pre-inflationary part
of the bounce, until the bouncing point, we shall now demonstrate
that for the specific model, the energy spectrum of the primordial
gravitational waves is overall amplified. The analysis of
primordial gravitational waves in the context of $f(R)$ gravity
was performed in detail in \cite{Odintsov:2021kup}, so let us
discuss in brief the essential features of the formalism here. For
a perturbed FRW metric, the Fourier transformation of the tensor
perturbation satisfies the following evolution equation,
\begin{equation}\label{fouriertransformationoftensorperturbation}
\frac{1}{a^3f_R}\frac{{\rm} d}{{\rm d} t}\left(a^3f_R\dot{h}(k)
\right)+\frac{k^2}{a^2}h(k)=0\, ,
\end{equation}
which can be cast in the following way,
\begin{equation}\label{mainevolutiondiffeqnfrgravity}
\ddot{h}(k)+\left(3+\alpha_M
\right)H\dot{h}(k)+\frac{k^2}{a^2}h(k)=0\, ,
\end{equation}
where the time-dependent parameter $\alpha_M$ for the $f(R)$
gravity case at hand is,
\begin{equation}\label{amfrgravity}
a_M=\frac{f_{RR}\dot{R}}{f_RH}\, .
\end{equation}
The study of primordial gravitational waves evolution can be
significantly simplified by adopting the WKB approach of Refs.
\cite{Nishizawa:2017nef,Arai:2017hxj}, which we now describe in
brief. We rewrite the evolution equation
(\ref{mainevolutiondiffeqnfrgravity}) of the tensor perturbations
as follows,
\begin{equation}\label{mainevolutiondiffeqnfrgravityconftime}
h''(k)+\left(2+a_M \right)\mathcal{H} h'(k)+k^2h(k)=0\, ,
\end{equation}
where the prime indicating this time differentiation with respect
to the conformal time $\tau$,  and $\mathcal{H}=\frac{a'}{a}$. The
WKB solution to the differential equation
(\ref{mainevolutiondiffeqnfrgravityconftime}) has the following
form,
\begin{equation}\label{mainsolutionwkb}
h=e^{-\mathcal{D}}h_{GR}\, ,
\end{equation}
with $h_{GR}$ is the GR waveform solution of the differential
equation (\ref{mainevolutiondiffeqnfrgravityconftime})
corresponding to $a_M=0$. Moreover, $\mathcal{D}$ is defined as
follows,
\begin{equation}\label{dform}
\mathcal{D}=\frac{1}{2}\int^{\tau}a_M\mathcal{H}{\rm
d}\tau_1=\frac{1}{2}\int_0^z\frac{a_M}{1+z'}{\rm d z'}\, .
\end{equation}
The GR waveform produces the following energy spectrum of the
primordial gravitational waves,
\begin{equation}
    \Omega_{\rm gw}(f)= \frac{k^2}{12H_0^2}\Delta_h^2(k),
    \label{GWspec}
\end{equation}
with $\Delta_h^2(k)$ being defined as
\cite{Boyle:2005se,Nishizawa:2017nef,Arai:2017hxj,Nunes:2018zot,Liu:2015psa,Zhao:2013bba,Odintsov:2021kup},
\begin{equation}\label{mainfunctionforgravityenergyspectrum}
    \Delta_h^2(k)=\Delta_h^{({\rm p})}(k)^{2}
    \left ( \frac{\Omega_m}{\Omega_\Lambda} \right )^2
    \left ( \frac{g_*(T_{\rm in})}{g_{*0}} \right )
    \left ( \frac{g_{*s0}}{g_{*s}(T_{\rm in})} \right )^{4/3} \nonumber  \left (\overline{ \frac{3j_1(k\tau_0)}{k\tau_0} } \right )^2
    T_1^2\left ( x_{\rm eq} \right )
    T_2^2\left ( x_R \right ),
\end{equation}
and note that the ``bar'' in the Bessel function term, denotes the
average taken over many periods. The details on the above
parameters can be found in
\cite{Boyle:2005se,Nishizawa:2017nef,Arai:2017hxj,Nunes:2018zot,Liu:2015psa,Zhao:2013bba,Odintsov:2021kup}.
Moreover, the term $\Delta_h^{({\rm p})}(k)^{2}$ denotes the
primordial tensor power spectrum generated during the inflationary
era, which is
\cite{Boyle:2005se,Nishizawa:2017nef,Arai:2017hxj,Nunes:2018zot,Liu:2015psa,Zhao:2013bba,Odintsov:2021kup},
\begin{equation}\label{primordialtensorpowerspectrum}
\Delta_h^{({\rm
p})}(k)^{2}=\mathcal{A}_T(k_{ref})\left(\frac{k}{k_{ref}}
\right)^{n_T}\, ,
\end{equation}
and note that this is evaluated at the CMB pivot scale
$k_{ref}=0.002$$\,$Mpc$^{-1}$. Moreover, $n_T$ is the inflationary
tensor spectral index and $\mathcal{A}_T(k_{ref})$ denotes the
primordial amplitude of the tensor perturbations, which can be
written in terms of the amplitude of the scalar perturbations
$\mathcal{P}_{\zeta}(k_{ref})$ as follows,
\begin{equation}\label{amplitudeoftensorperturbations}
\mathcal{A}_T(k_{ref})=r\mathcal{P}_{\zeta}(k_{ref})\, ,
\end{equation}
where $r$ is the tensor-to-scalar ratio, hence we finally have,
\begin{equation}\label{primordialtensorspectrum}
\Delta_h^{({\rm
p})}(k)^{2}=r\mathcal{P}_{\zeta}(k_{ref})\left(\frac{k}{k_{ref}}
\right)^{n_T}\, .
\end{equation}
Combining the above, the energy spectrum for $f(R)$ gravity is,
\begin{align}
\label{GWspecfR}
    &\Omega_{\rm gw}(f)=e^{-2\mathcal{D}}\times \frac{k^2}{12H_0^2}r\mathcal{P}_{\zeta}(k_{ref})\left(\frac{k}{k_{ref}}
\right)^{n_T} \left ( \frac{\Omega_m}{\Omega_\Lambda} \right )^2
    \left ( \frac{g_*(T_{\rm in})}{g_{*0}} \right )
    \left ( \frac{g_{*s0}}{g_{*s}(T_{\rm in})} \right )^{4/3} \nonumber  \left (\overline{ \frac{3j_1(k\tau_0)}{k\tau_0} } \right )^2
    T_1^2\left ( x_{\rm eq} \right )
    T_2^2\left ( x_R \right )\, ,
\end{align}
and recall that $\mathcal{D}$ is defined in Eq. (\ref{dform}).
Thus what is required in order to reveal the effect of the $f(R)$
gravity, is to calculate the parameter $\mathcal{D}$ from $z=0$ up
to redshift it is required. Now we are interested in ultrahigh
redshifts corresponding to pre-inflationary times. In order to
quantify our study, the calculation of the parameter $\mathcal{D}$
can be divided in two redshift eras, the era from present day up
to redshifts corresponding to the end of inflation, and for
redshifts corresponding to the end of inflation up to the bouncing
point. For the calculation of the redshift, it is useful to have
the relation between the redshift and the temperature of the
Universe, which is approximately \cite{Garcia-Bellido:1999qrp},
\begin{equation}\label{redshiftvstemperature}
T=T_0(1+z)\, ,
\end{equation}
with $T_0$ being the present day temperature, which is 3 Kelvin,
or $T_0=2.58651\times 10^{-4}$eV. With this relation, one can
easily find the required redshifts for the calculation of the
parameter $\mathcal{D}$ and subsequently the calculation of the
energy power spectrum of the primordial gravitational waves. The
parameter $\mathcal{D}$ will be numerically evaluated in the
aforementioned redshift eras, and it is equal to,
\begin{equation}\label{dformexplicitcalculation}
\mathcal{D}=\frac{1}{2}\left(\int_0^{z_f}\frac{a_{M_1}}{1+z'}{\rm
d z'}+\int_{z_i}^{z_b}\frac{a_{M_2}}{1+z'}{\rm d z'}\right)\, .
\end{equation}
where $z_f$ is the redshift of the Universe at the end of
inflation, while $z_i$ and $z_b$ denote the redshifts at the
beginning of inflation and at the bouncing point. One may think
that we can calculate the effects of the era before the bouncing
point, however this is not possible. This is due to the fact that
for the numerical calculation of the parameter $\mathcal{D}$ from
present time, one uses the relation $1+z=\frac{1}{a}$, thus the
scale factor at present time is unity and as the redshift
increases, the scale factor decreases (we have set the present
time scale factor equal to unity in order for the comoving and
physical wavelengths and relevant quantities to coincide). The
scale factor decreases from present time until the bouncing point,
where it reaches its minimum value, and beyond the bouncing point,
the scale factor increases again. Therefore the era before the
bouncing point is not reachable by the present time era using
$1+z=\frac{1}{a}$, it is an inaccessible era numerically. Note
that the parameters $a_{M_1}$ and $a_{M_1}$ appearing in Eq.
(\ref{dformexplicitcalculation}) both correspond to relation
(\ref{amfrgravity}), but for different forms of the $f(R)$
gravity. Specifically, for the era between present time and the
end of inflation, the $f(R)$ gravity is assumed to be described
by,
\begin{equation}\label{starobinsky}
f(R)=R+\frac{1}{M^2}R^2-\gamma \Lambda
\Big{(}\frac{R}{3m_s^2}\Big{)}^{\delta}\, ,
\end{equation}
with $m_s$ in Eq. (\ref{starobinsky}) being
$m_s^2=\frac{\kappa^2\rho_m^{(0)}}{3}$, and $\rho_m^{(0)}$ is the
energy density of cold dark matter today. The model
(\ref{starobinsky}) is basically the post-bouncing point $f(R)$
gravity plus a power-law correction term which does not affect the
inflationary era, since $0<\delta<1$, but does seriously affect
the late-time era. The complete lat-time study of the model
(\ref{starobinsky}) was performed in Ref. \cite{Odintsov:2021kup}
and it was shown that a viable dark energy era can be obtained and
also the calculation of the the integral
$\int_0^{z_f}\frac{a_{M_1}}{1+z'}{\rm d z'}$ yields the result
$-0.004$, thus the amplification for this era is $\sim
e^{0.00451304}=1.00452$. For the calculation of the term
$\int_{z_i}^{z_b}\frac{a_{M_2}}{1+z'}{\rm d z'}$, which
corresponds to the pre-inflationary and post-bouncing point era,
the $f(R)$ gravity is basically that of Eq.
(\ref{frformprevcubic}). As it proves, the values of the
parameters $\mathcal{C}_{3,4}$ and $H_b$ do not significantly
affect the results, so we shall assume that $H_b\sim
\mathcal{O}(M_p)$ and $\mathcal{C}_{3,4}\sim \mathcal{O}(M_p^2)$.
Regarding the redshifts $z_i$ and $z_b$ we shall calculate them
using Eq. (\ref{redshiftvstemperature}). Specifically, for the
redshift $z_i$ which corresponds to the beginning of inflation, we
shall take the corresponding temperature to be the temperature of
inflation, so $T_{inf}\sim 10^{15}$GeV, hence by using Eq.
(\ref{redshiftvstemperature}) $z_i$ is $z_i=3.86\times 10^{27}$.
Regarding the bouncing point temperature, the bouncing point
basically describes the most dense state in the Universe, so
string theory effects should in principle be present. The
classical description should also in principle fail, hence it
would be inaccessible from us in the way we described above.
However, for the shake of simplicity of the argument, we shall
assume that we can reach it, because it proves that the final
redshift does not affect significantly the calculation of the
integral, at least it does not affect the result in a major and
reportable way. Hence let us assume that the temperature of the
bouncing point is the Planck temperature $T_b\sim 10^{19}$GeV
hence $z_b\simeq 3.86\times 10^{31}$. Now the numerical
calculation of the integral is easy and it yields
$\int_{z_i}^{z_b}\frac{a_{M_2}}{1+z'}{\rm d z'}=-0.0338031$. Thus
the overall amplification effect of the $f(R)$ gravity is
$e^{-2\mathcal{D}}\sim e^{0.03831}=1.03906$. In Fig.
\ref{plotfinalfrpure} we present the  $h^2$-scaled gravitational
wave energy spectrum for pure $f(R)$ gravity and we also plot the
sensitivity curves of the most important planned future
gravitational waves experiments. The $f(R)$ gravity power spectrum
is significantly below the sensitivity curves of the planned
future experiments. Hence, although the pre-inflationary era leads
to an amplification of the overall energy power spectrum, the
signal is still undetectable. Thus if in future gravitational wave
experiments some signal is detected in the high frequency range,
this in principle will exclude $f(R)$ gravity models and all the
slow-roll scalar field models, unless some exotic reheating era
phenomenon takes place. Of course, the present result is model
dependent and in principle some specific $f(R)$ gravity model may
eventually yield a significant amplification, however it seems
that $f(R)$ gravity in the presence of a scalar field might lead
to a significant amplification of the primordial gravitational
wave power spectrum. In Ref. \cite{Odintsov:2021kup} we presented
an example of that sort, however we used a Chern-Simons, however
we have reportable theoretical evidence that even a simpler
$f(R,\phi)$ theory can yield an amplified gravitational wave power
spectrum. We shall report on this issue soon. In conclusion,
amplification in $f(R)$ gravity will occur if in addition to the
higher derivative terms, a scalar field is present.

\section{Conclusions}

In this paper we studied an alternative evolution for our
primordial Universe, which mainly consists of a non-singular
bounce prior to inflation. Specifically, the Universe contracts
prior to inflation until it reaches a minimum magnitude, where it
bounces off and thereafter after a short period of time a quasi-de
Sitter era commences. After the end of the inflationary era, the
matter and radiation perfect fluids cannot be ignored, so the
radiation era commences and the Universe decelerates as in
ordinary post-inflationary scenarios. We investigated how these
different evolution patches can be realized by vacuum $f(R)$
gravity, and we found the pre-inflationary description of the
model, and in addition the inflationary quasi-de Sitter era phase,
which is described by an $R^2$ gravity. The $R^2$ model in vacuum
has unstable de Sitter vacua in the phase space
\cite{Odintsov:2017tbc} which allow a natural exit from inflation.
The non-singular pre-inflationary bounce has some interesting
features, for example it respects the scale factor duality during
the pre-inflationary and post-inflationary era. In addition,
motivated by string theory pre-inflationary scenarios which lead
to an amplified gravitational wave energy spectrum, we examined
whether the pre-inflationary bounce effects may lead to an
amplification of the $f(R)$ gravity gravitational wave energy
spectrum. As we showed in detail, indeed an overall amplification
occurs for the $f(R)$ gravity theory, but the amplification is
very small, and in effect, the resulting energy spectrum lies
below the sensitivity curves of most future interferometer
experiments. Although the result is somewhat model dependent, it
seems that a pattern emerges for $f(R)$ gravity theories, in
common ground with single scalar field theories, which indicates
that these theories are not detectable if they are considered by
themselves. Hence if a future interferometer verifies the
existence of a stochastic gravitational wave background, this will
indicate either that the physics are not controlled by single
scalar field theory or $f(R)$ gravity, or that some exotic
scenario for the reheating era takes place. Even so, one has to
try hard to amplify the spectrum of $f(R)$ gravity and single
scalar field theory by themselves. However, note that our
conclusion about small amplification is valid only for  specific
pre-inflationary bounce under consideration. However, we have
strong theoretical evidence which we will report soon, that a
combination of higher curvature $f(R)$ gravity and non-minimally
coupled scalar fields yields a considerable amplification of the
gravitational wave energy spectrum, which eventually will be
measurable by most, or even all the high frequency interferometer
experiments. One example of this sort was reported in
\cite{Odintsov:2021kup}, but in the presence of exotic parity
violating terms of Chern-Simons type. As we show in a future work,
it is possible from a combined $f(R,\phi)$ to obtain a significant
amplification for the energy spectrum of the primordial
gravitational waves.

\section*{Acknowledgments}

This work was supported by MINECO (Spain), project
PID2019-104397GB-I00 (S.D.O).

\end{document}